\documentclass[twocolumn,showpacs,aps]{revtex4}
\usepackage{amsmath}
\usepackage{amssymb}

\newcommand{\tr}{{\rm Tr}}
\newcommand{\comm}[2]{[#1,#2]}

\begin{document}

\title{
The Second Law of Steady State Thermodynamics for Nonequilibrium
Quantum Dynamics
}
\author{Satoshi Yukawa}
\email{yukawa@ap.t.u-tokyo.ac.jp}
\affiliation{Department of Applied Physics, School of Engineering,
  The University of Tokyo,\\ Hongo, Bunkyo-ku, Tokyo 113-8656,
  Japan}
\date{\today}

\begin{abstract}
  The second law of ordinary thermodynamics and the second law of
  steady state thermodynamics, as proposed by Oono and Paniconi, are
  investigated from the microscopic point of view for the open quantum
  system. Based on the H-theorem of Lindblad, we show that both second
  laws are consistent with the quantum dynamics generated by the
  completely positive map.  In addition, microscopic expressions of
  entropy production and ``housekeeping heat'' are obtained for some
  classes of dynamical quantum systems.
\end{abstract}
\maketitle

Understanding the properties of nonequilibrium states of quantum systems
may become important for developing quantum devices which perform
efficiently in nonequilibrium situations; It is known, for
instance, that quantum ratchet systems only work in competitive
situations between thermal relaxation and external
driving\cite{YKTM97,RGH97,TKYM98}. A theoretical framework for
considering the general properties of nonequilibrium states, however,
is still under development.

In this letter we focus our attention on thermodynamic properties of
steady states in nonequilibrium quantum dynamics, especially the
second law of steady state thermodynamics. The steady state
thermodynamics (SST) proposed by Oono and
Paniconi\cite{OP98} is a phenomenological thermodynamic framework for
the steady state instead of thermal equilibrium.  The second law of
SST is not simple because the steady state produces heat constantly;
In such a situation the total heat is divergent.  Therefore the
meaning of heat in the second law of SST is ambiguous.  Oono and
Paniconi introduced a concept of ``excess heat'' to resolve this
confusing situation.  The excess heat is defined by subtracting 
``housekeeping heat'' from the total heat; The housekeeping heat
is defined as the energy 
preventing
steady states from thermal
relaxation.  Then the excess heat for the steady state becomes finite
and we can discuss the second law with this excess heat.  There are
still ambiguities; How do we determine other thermodynamic quantities
operationally? What are microscopic expressions of such quantities?
In this letter we consider microscopic quantum descriptions;
Especially we treat some classes of small  open  quantum systems and
thermodynamic processes for such systems. As a consequence we discuss
the second law of SST for nonequilibrium quantum dynamics of small
open quantum systems.

Our strategy is as follows: First we clarify how the equilibrium
thermodynamics is consistent with the nonequilibrium quantum dynamics.
Next we extend this result to the nonequilibrium steady state
case. Mathematically this operation has no ambiguity for some limiting
class of the system.  Finally we discuss the validity of our expression.

The present quantum system we consider consists of three parts; One is
a main system with a time-varying field. This part is operated from the
outside. The time-varying field acts like, for instance, a piston in a
cylinder. Our interest is concentrated on this part. We write its
Hamiltonian as $H(t)$ or $H_\lambda$ explicitly representing the
time-dependent field $\lambda(t)$.  Here we assume that the system is
sufficiently small. 
The second part of the whole system is a heat bath. It is so large that
it is always in thermal equilibrium with a definite temperature (we
denote its inverse temperature as $\beta$).  The last part is an
interaction between the first two. The coupling constant $\gamma$ governs
the relaxation time of the system.  
In the present study we assume that the coupling is sufficiently
weak. This assumption ensures the independency of the small system from
the heat bath. Then we can discuss the thermodynamic properties of the
small system. 
It should be noted that a
temperature of the small system cannot be defined in its nonequilibrium
state.  We only mention its temperature in the thermal equilibrium state
by the corresponding heat bath temperature $\beta$.

A thermodynamic process for the present study 
is described in the following way; At the initial time, a density
operator of the small system is taken to be in thermal equilibrium.
From this state we operate the field $\lambda$ from $0$ to $1$
(these values can be chosen arbitrary without loss of generality) in
the finite time interval $T$. 
The value of $T$ controls a property of the dynamics; 
If $T$ is sufficiently shorter than the characteristic time determined
by the minimal energy gap at the pseudo crossing of energy levels,  the
process may become nonadiabatic\footnote{ The condition that the
  non-adiabatic transition appears is discussed in a text book, for
  instance, written by A. Messia: \textit{quantum mechanics}, (Dover,
  New York, 1999)} 
in the quantum mechanical sense.  
If $T$ is sufficiently larger than the above time scale but much smaller
than the thermal relaxation time ($\sim O(\gamma^{-2})$), the adiabatic
dynamics may be actualized. 
If $T$ is much larger than all the above mentioned time
scales, the dynamics becomes a quasi-equilibrium. 
Then the system is always in the thermal equilibrium state
corresponding to the Hamiltonian $H_\lambda$.

The microscopic dynamics of the system we use here
is obtained by a projection method\cite{KTH91}.
After applying the projection method and weak coupling expansion, 
we obtain the equation of motion
for the density operator $\rho(t)$ of the system as follows:
\begin{equation}
\dfrac{\partial \rho(t)}{\partial t} = - \dfrac{i}{\hbar} 
\comm{H(t)}{\rho(t)} - \gamma^2 \Gamma_{H(t)}( \rho(t))
\enspace .
\label{eq:motion}
\end{equation}
The first term represents pure quantum dynamics. The second term is a
thermal relaxation term; If the Hamiltonian is independent of time,
the system relaxes into the thermal equilibrium state.  The result of
this letter is, however, independent of the details of dynamics.  Only
the (1) existence of the equilibrium or nonequilibrium steady states
and the (2) special features of the dynamics mentioned in latter part
is important.

Let us define thermodynamic quantities in the present situation.
First we define the internal energy. It can be regarded as the
expectation value of the Hamiltonian under the assumption that the
system is small. In the present thermodynamic process we can write a
total change of the internal energy $\Delta E$ as
\begin{equation*}
\Delta E \equiv \int_0^1 \dfrac{d\lambda}{\dot{\lambda}} 
\frac{\partial \langle H_\lambda \rangle_t }{\partial t}
= \langle H_1 \rangle_T  - \langle H_0 \rangle_0
\enspace ,
\end{equation*}
where $\dot{\lambda}$ is the time derivative of $\lambda$ and $\langle \dots \rangle_t$
represents $\tr( \dots \rho(t))$.

The total work $W$ done by the external field during the process is
defined as
\[
W \equiv \int_0^1 \dfrac{d\lambda}{\dot{\lambda}}  
\langle \frac{\partial H_\lambda }{\partial t } \rangle_t
\enspace ,
\]
that is, the work is the summation of the average value of the change of
the Hamiltonian with respect to $\lambda$\cite{Le78,PW78}.
This definition of work is dependent on the thermodynamic process
in contrast to the change of the internal energy.

Using $\Delta E$ and $W$, we can define the ``heat'' $\Delta Q$ in this
context as 
\[
\Delta Q \equiv \Delta E - W
\enspace .
\]
This quantity represents the net energy coming from the heat bath.
It is written by microscopic variables as
\[
\Delta Q = - \frac{1}{\beta} \int_0^1 \dfrac{d\lambda}{\dot{\lambda}}
   \tr \left( \frac{\partial \rho(t)}{\partial t} \ln \rho^{eq}(t)  \right) 
\enspace ,
\]
where $\rho^{eq}(t)=\exp\left( -\beta H(t) \right) /\tr \exp\left( -\beta H(t)
\right)$ is the thermal equilibrium density matrix.

The definition of the entropy is not clear, because the entropy is
a concept in an equilibrium state, not a dynamical state.
Therefore we need some definitions of entropy. Here we use 
the von Neumann entropy:
\begin{equation}
S(t) = - \tr \rho(t) \ln \rho(t)
\enspace.
\label{eq:vonNeumann}
\end{equation}
This definition is consistent with the equilibrium statistical mechanics.
However it is not trivial whether this is valid for nonequilibrium
dynamical states; 
For instance, the von Neumann entropy neither increases nor decreases
for any nonadiabatic process represented by unitary evolutions,
even though classical thermodynamics states that the entropy
increases. In order to treat 
this paradoxical situation, one frequently takes the partial trace for
the microscopic degrees of freedom\cite{Zu74}. This process is called a
``coarse-graining''. 
In the present case, the explicit coarse-graining is not
suitable because the system has a few degrees of freedom; It is not
clear which degrees of freedom are traced out. 
Instead, because of the non-unitary evolution, 
the effect of the coarse-graining is introduced naturally in 
Eq.~(\ref{eq:vonNeumann}).
We obtain the change of entropy during the process as 
\begin{equation*}
\Delta S = \int_0^1 \dfrac{d\lambda}{\dot{\lambda}} 
\frac{\partial  S(t)}{\partial t} = S(T) - S(0)
\enspace .
\end{equation*}

Now we can write the energy conservation as
\begin{equation}
\frac{1}{\beta} \left( \Delta S - \Delta \epsilon \right) =  \Delta Q
\enspace ,
\label{eq:first}
\end{equation}
where $\Delta \epsilon$ is calculated as 
\begin{equation}
\Delta \epsilon =  \int_0^1 \dfrac{d\lambda}{\dot{\lambda}} 
\tr 
\dfrac{\partial \rho}{\partial t} 
\left( 
  \ln \rho^{eq}(t) - \ln \rho(t)
\right)
\enspace .
\label{eq:excess}
\end{equation}
Equation~(\ref{eq:first}) is the ordinary first law of
thermodynamics without $\Delta \epsilon$.  We can easily show that
$\Delta \epsilon$ becomes zero in the quasi-equilibrium or the non-adiabatic
limits. Therefore nonzero values of $\Delta \epsilon$ represent
the irreversibility of the dynamics. We call $\Delta \epsilon$ 
``entropy production''.

If the non-negativity  of $\Delta \epsilon$ can be proved, we obtain the
following inequality:
\begin{equation}
  \frac{1}{\beta} \Delta S \geq \Delta Q
  \enspace . 
  \label{ieq:clausius}
\end{equation}
This form is quite similar to the original Clausius inequality. But
there is an essential difference. 
In the original one the entropy difference is 
is one of two thermal equilibrium values. 
By contrast $\Delta S$ in Eq.~(\ref{ieq:clausius}) is a dynamical value
dependent on the process connecting two states.  
In addition there is no necessity that these two states are
equilibrium states.

Let us consider the non-negativity of the entropy production rate $\epsilon
\equiv \tr \left( \dot{\rho} \left( \ln \rho^{eq}(t) - \ln \rho(t)
\right) \right)$ instead of the entropy production. 
Roughly speaking, this term can be non-negative;
If the current density operator is ``larger'' (``smaller'') than one
corresponding to the equilibrium state, the
part $\ln \rho^{eq}(t) - \ln \rho(t)$ is ``less'' (``grater'') than $0$.
 If the system has normal thermodynamic stability, we can
expect that the derivative  $\dot{\rho}$ is negative (positive) at the same time.  
In this way, $\epsilon$ becomes non-negative.

For a more precise discussion, we define the following quantity:
\[
d(\rho_1, \rho_2) \equiv \tr \rho_1 (\ln \rho_1 - \ln \rho_2)
\]
for two arbitrary density operators $\rho_1, \rho_2$. This quantity is
known as a quantum relative entropy or a quantum Kullback
divergence\cite{AN00}, which is a kind of distance between $\rho_1$ and $\rho_2$.
We can prove that the above quantity is always
non-negative. If and only if $\rho_1 = \rho_2$, $d(\rho_1,\rho_2)$
vanishes. Using $d$, we naively rewrite $\epsilon$ as
\begin{equation*}
  \epsilon = \lim_{dt \to 0} \frac{1}{dt} \left[ d(\rho(t), \rho^{eq}(t)) - 
  d(\rho(t+dt), \rho^{eq}(t)) \right] \enspace . 
\end{equation*} 
Therefore the non-negativity condition of $\epsilon$ becomes
\begin{equation}
d(\rho(t+dt), \rho^{eq}(t)) \leq d(\rho(t),\rho^{eq}(t))
\label{ieq:hthe}
\end{equation}
for sufficiently small positive $dt ( \ll T)$.

In 1975, Lindblad proved that the above inequality for more general
cases\cite{Li75}. His statement is as follows: For a completely
positive map $K$, the inequality
\begin{equation}
  \label{ieq:Lindblad}
  d(K \rho_1, K \rho_2)  \leq d(\rho_1,\rho_2)
\end{equation}
holds\footnote{Classical version of this inequality can be found in
  the text book of information theory: T. M. Cover and J. A. Thomas,
  \textit{Elements of Information Theory}, (Wiley, New York, 1991).}.  
Here $\rho_1$ and $\rho_2$ are some density matrices of the
system. The completely positive map is a trace-preserving linear map
keeping the positivity of the density matrices. 
The statement ``completely'' means that the dynamics of the system
cannot be influenced by the outside and cannot influence the outside.
For example, a simple unitary evolution is a completely positive map. 
A projected map of the unitary evolution is also a completely positive
map.

Here we sketch the outlines of his proof. It is based on the following
two things: (1) the joint convexity of the relative
entropy\cite{Li73}, 
\begin{equation}
\label{ieq:jconv}
d(\sum_i \lambda_i \rho_i , \sum_i \lambda_i \sigma_i) \leq \sum_i \lambda_i d(\rho_i, \sigma_i)
\enspace, 
\end{equation}
where $\rho_i, \sigma_i$ are two density matrices and $\lambda_i$ are
non-negative numbers restricted by $\sum_i \lambda_i = 1$, and (2) the
decomposition of the completely positive map $K$ by the unitary
evolution $U$;
\begin{equation}
\label{eq:dec}
K \rho = \sum_{U} K(U) U^\dagger \rho U 
\enspace , 
\end{equation}
where $\sum_{U} K(U) = 1, \,\, K(U)  > 0$. 
If we substitute Eq.~(\ref{eq:dec}) into the left hand side of
inequality~(\ref{ieq:Lindblad}) and use the properties of the coefficients
$K(U) $ and the joint convexity of $d$, we get Lindblad's
H-theorem.

In the present case our dynamics is written as a completely
positive map\cite{Li76,GKS76}. If we choose $\rho_1$ as $\rho(t)$, $\rho_2$ as 
$\rho^{eq}(t)$ and $K$ as the infinitesimal time-evolution operator of
the system $\Phi_t^{dt}$, 
we get 
\[
d(\Phi_t^{dt} \rho(t) , \Phi_t^{dt} \rho^{eq}(t)) 
\leq 
d(\rho(t), \rho^{eq}(t)) \enspace .
\]
Since the instantaneous thermal equilibrium state $\rho^{eq}(t)$ is the
invariant state of the map at time $t$, we obtain the inequality
(\ref{ieq:hthe}). 
Therefore the inequality~(\ref{ieq:clausius}) holds.

We emphasize again that the dynamical Clausius
inequality~(\ref{ieq:clausius}) is dependent on a process  which changes
the parameter $\lambda$. If we want to discuss the second law of
thermodynamics, we have to compare the present entropy change $\Delta
S$ with the quasi-equilibrium change of the entropy $\Delta
S_{qs}$. The latter is only dependent on the two equilibrium
distributions of the system with $\lambda = 0, 1$. 
To get the usual second law, we start from the thermal equilibrium
state with $\lambda = 0 $ and leave the system after the switching
process with $\lambda = 1$ fixed. After relaxation, the quantities
$\Delta S$ and $\Delta E$ converge into quasi-equilibrium differences
$\Delta S_{qs}$ and $\Delta E_{qs}$. But the work $W$ does not change. 
Then we get the Clausius inequality:
\[
 \dfrac{1}{\beta}  \Delta S_{qs} \geq \Delta E_{qs} - W \enspace .
\]

Let us consider the second law of steady state thermodynamics based on
the knowledge of the previous consideration; Key points are the
existence of the fixed point density matrix for the time-evolution map
and Lindblad's H-theorem. Mathematically it is not important whether
the fixed point is a thermal equilibrium state or not. 
If there is a nonequilibrium steady state described by $\rho^{ss}(t)$,
the system may relax into this state, not into the 
equilibrium state $\rho^{eq}(t)$.
In such case, we can obtain a second-law analogue for the steady state
by substituting $\rho^{eq}(t)$ with $\rho^{ss}(t)$, which is generally
dependent on time. 
Then we obtain the expression of the entropy production as 
\begin{multline}
  \Delta \epsilon = \int_0^1 \dfrac{d\lambda}{\dot{\lambda}} 
   \tr \left( \dfrac{\partial
      \rho}{\partial t} \left( \ln \rho^{ss}(t) - \ln \rho(t) \right)
  \right)\\
  + \int_0^1 \dfrac{d\lambda}{\dot{\lambda}}  \tr
  \left( \dfrac{\partial \rho}{\partial t} \left( \ln \rho^{eq}(t) -
      \ln \rho^{ss}(t) \right) \right) \enspace .
\end{multline}
We define an entropy production for the steady state as
\begin{equation}
  \Delta \epsilon^{ss} \equiv \int_0^1\dfrac{ d\lambda}{\dot{\lambda}}  \tr 
  \left(
    \dfrac{\partial \rho}{\partial t} \left(
      \ln \rho^{ss}(t) - \ln \rho(t)
    \right)
  \right)
\end{equation}
and define the difference of the ``housekeeping'' heat during the
process as 
\begin{equation}
\Delta Q_{hk} \equiv \frac{1}{\beta} \int_0^1\dfrac{ d\lambda }{\dot{\lambda}}
 \tr
\left(\dfrac{\partial \rho}{\partial t}
  \left(\ln \rho^{eq}(t) - \ln \rho^{ss}(t)\right)
\right)
\enspace .
\end{equation}
The housekeeping heat is consumed to keep the system in the steady
state, and prevent it form thermal relaxation. 
In the quasi-steady limit, the density operator $\rho(t)$ always follows
$\rho^{ss}(t)$. Then $\Delta Q_{hk} $ can be regarded as the entropy 
production for the equilibrium state which is non-negative.

The corresponding first law is obtained as follows:
\begin{equation}
\dfrac{1}{\beta} \left(\Delta S - \Delta \epsilon^{ss}\right) =
\Delta Q + \Delta Q_{hk}
\enspace .
\end{equation}
In the present case we can expect that the entropy production for the
steady state is non-negative. Therefore the following new inequality  
\begin{equation}
\dfrac{1}{\beta} \Delta S \geq
\Delta Q + \Delta Q_{hk}
\label{ieq:2ndsst}
\end{equation}
holds. This inequality is the second law of
thermodynamics for the steady state in our present situation.

Next we will look at other expressions for the  housekeeping heat and
consider its properties.  
We denote the steady state as $\rho^{ss}(t) = \exp ( -\beta (H(t) +
\phi(t))) / Z^{ss}(t)$ with the steady state partition function  $Z^{ss}
(t) = \tr\exp ( -\beta (H(t) + \phi(t))) $. $\phi(t)$ represents the nonequilibrium part of
Hamiltonian. Then the housekeeping heat can be written as a
fluctuation form:
\begin{multline}
\Delta Q_{hk} = \beta \int_0^1 \dfrac{d\lambda }{\dot{\lambda}}
\left\{
  \left\langle \dfrac{\partial
    H(t)}{\partial t} + \dfrac{\partial \phi(t)}{\partial t}
  \right\rangle_t^{ss}  \langle \phi(t) \rangle_t^{ss} \right.\\
  - \left.\left\langle \left(\dfrac{\partial
    H(t)}{\partial t} + \dfrac{\partial \phi(t)}{\partial t}
  \right)
  \phi(t)
  \right\rangle_t^{ss}
\right\}
\enspace ,
\end{multline}
where $\langle \dots \rangle_t^{ss}$ represents $\tr ( \dots \rho^{ss}(t))$,
or more simply 
\[
\Delta Q_{hk} = \int_0^1 \dfrac{d\lambda}{\dot{\lambda}} 
\tr \left( \dfrac{\partial \rho^{ss}(t)}{\partial t} \phi(t) \right)
\enspace .
\]
In the quasi-steady cyclic process, the  differences of entropy, energy, and
entropy production for the steady state vanish. In this case, 
normal work and ``$\phi$-work'' is balanced, that is, 
\[
\int_0^1 \dfrac{d\lambda}{\dot{\lambda}} 
\left\langle 
  \dfrac{\partial H}{\partial t} 
\right\rangle_t^{ss}
= 
\int_0^1 \dfrac{ d\lambda}{\dot{\lambda}} 
\left\langle 
  \dfrac{\partial \phi}{\partial t} 
\right\rangle_t^{ss}
\enspace .
\]
This condition imposes some restrictions on the determination of
the nonequilibrium part of $\phi(t)$.

It is of note that $\Delta Q_{hk}$ is the difference of housekeeping heat for the
dynamical process. If the density matrix does not depend on time, $\Delta
Q_{hk}$ as defined above is strictly zero.  Therefore we should consider that
the present definition of housekeeping heat describes the dynamical
part of total housekeeping heat; The remaining housekeeping heat cannot
be calculated with the above definition, because it is stationary with
respect to time. This is not surprising, since our system is
dynamical.  We assume that the work done by outside is mechanical 
and dependent on time.  Using this type of work, we cannot
realize a stationary nonequilibrium steady state accompanied by
housekeeping heat. We only achieve making time-dependent
nonequilibrium steady states like a limit cycle or stationary
nonequilibrium states without housekeeping heat.

To treat the stationary nonequilibrium steady state, we have to take into
account the  nonequilibrium boundary conditions such as thermal
gradients or chemical potential gradients.  In the present situation,
however, this cannot be described by our definition of work. Therefore
we need other expressions of thermodynamic quantities. If we obtain
these expressions, the corresponding first and second laws may be
obtained. But this is outside our scope to deal with that matter in this
letter.

To summarize, we have shown that the second law of thermodynamics is
consistent with microscopic quantum dynamics of open systems. In
addition, for the steady state thermodynamics, it is also true that
 the second law in the steady-state-based thermodynamic process of
the open quantum system exists. In this letter we have studied the
second laws for the limited classes of open quantum system and
dynamics.  But similar discussions for some other systems, for
instance, one with nonequilibrium boundary conditions, may be
possible. Especially for classical Markov systems, the present
argument is directly applicable with slight
modifications\footnote{S. Sasa, private communication }. For the
classical Langevin system, the second law of SST also
holds\cite{HS01}. 

The author is grateful to N. Ito, M. Kikuchi, S. Miyashita, S. Sasa,
K. Sekimoto, H. Tasaki for valuable discussions.
The author also thanks H.-G. Matuttis for critical reading of the
manuscript.


\end{document}